\documentstyle[12pt]{article}

\begin{document}

\setcounter{page}{1}

\pagestyle{plain} \vspace{1cm}
\begin{center}
\Large{\bf One more step toward the noncommutative brane inflation}\\
\small \vspace{1cm} {\bf Kourosh Nozari$^{*}$} \quad and \quad {\bf
Siamak Akhshabi$^{\dag}$}\\ \vspace{0.5cm} {\it Department of
Physics, Faculty of Basic
Sciences,\\
University of Mazandaran,\\
P. O. Box 47416-95447, Babolsar, IRAN\\
\vspace{0.3cm}
$^{*}$knozari@umz.ac.ir\\
$^{\dag}$ s.akhshabi@umz.ac.ir}
\end{center}
\vspace{1.5cm}
\begin{abstract}
Recently a new approach to inflation proposal has been constructed
via the smeared coherent state picture of spacetime
noncommutativity. Here we generalize this viewpoint to a
Randall-Sundrum II braneworld scenario. This model realizes an
inflationary, bouncing solution without recourse to any axillary
scalar or vector fields. There is no initial singularity and the
model has the potential to produce scale invariant spectrum of
scalar perturbations.
\end{abstract}
\section{Introduction}
Spacetime non-commutativity can be achieved naturally on certain
backgrounds of string theory [1,2]. Existence of a fundamental
minimal length of the order of the Planck length and spacetime
non-commutativity are naturally related in these theories [3]. In
this viewpoint, description of the spacetime as a smooth commutative
manifold becomes therefore a mathematical assumption no more
justified by physics. It is then natural to relax this assumption
and conceive a more general noncommutative spacetime, where
uncertainty relations and spacetime discretization naturally arise.
Noncommutativity is the central mathematical concept expressing
uncertainty in quantum mechanics, where it applies to any pair of
conjugate variables, such as position and momentum. One can just as
easily imagine that position measurements might fail to commute and
describe this using noncommutativity of the coordinates. The
noncommutativity of spacetime can be encoded in the commutator [1]
\begin{equation}
[\hat{x}^i,\hat{x}^j]=i\theta^{ij}
\end{equation}
where $\theta^{ij}$ is a real, antisymmetric and constant tensor,
which determines the fundamental cell discretization of spacetime
much in the same way as the Planck constant $\hbar$ discretizes the
phase space. In $d = 4$, it is possible by a choice of coordinates
to bring some $\theta^{ij}$s to the following form
\begin{displaymath}
\theta^{ij}= \left( \begin{array}{cccc}
0 & \theta & 0 & 0  \\
-\theta & 0 & \theta & 0 \\
0 & -\theta & 0 & \theta \\
0 & 0 & -\theta & 0
\end{array} \right)
\end{displaymath}
This was motivated by the need to control the divergences showing up
in theories such as quantum electrodynamics. Here $\sqrt{\theta}$ is
the fundamental minimal length ( order of magnitude of
$\sqrt{\theta}$ can be found in Ref. [3]). This noncommutativity
leads to the modification of Heisenberg uncertainty relation in such
a way that prevents one from measuring positions to better
accuracies than the Planck length.

It has been shown that noncommutativity eliminates point-like
structures in the favor of smeared objects in flat spacetime. As
Nicolini {\it et al.}\, have shown [4] ( see also [5] for extensions
), the effect of smearing is mathematically implemented as a
substitution rule: position Dirac-delta function is replaced
everywhere with a Gaussian distribution of minimal width
$\sqrt{\theta}$. In this framework, they have chosen the mass
density of a static, spherically symmetric, smeared, particle-like
gravitational source as follows
\begin{equation}
\rho_\theta(r)=\frac{M}{(2\pi\theta)^{\frac{3}{2}}}\exp(-\frac{r^2}{4\theta}).
\end{equation}
As they have indicated, the particle mass $M$, instead of being
perfectly localized at a point, is diffused throughout a region of
linear size $\sqrt{\theta}$. This is due to the intrinsic
uncertainty as has been shown in the coordinate commutators (1).\\

Recently, a new approach to the issue of inflation in the framework
of Nicolini {\it et al.}\,  coherent states viewpoint of
noncommutativity has been reported by Rinaldi [6]. In this model,
the intrinsic noncommutative structure of spacetime is responsible
for a violation of the dominant energy condition near the initial
singularity, which induces a bounce. The following expansion is
quasi-exponential and it does not require any {\it ad hoc} scalar
field. Here we are going to investigate the effects of the spacetime
non-commutativity on the inflationary dynamics in the
Randall-Sundrum II braneworld scenario. We use the Nicolini {\it et
al.} coherent state approach encoded in the smeared picture defined
in (2). Some other studies of the non-commutative inflation with
different approaches can be found in Ref. [7].

\section{Noncommutative brane inflation}
We begin with the Randall-Sundrum II (RS II) geometry. In this
setup, there is a single positive tension brane embedded in an
infinite bulk [8]. The Friedmann equation governing the evolution of
the brane in this scenario is given as follows ( see for instance
[9])
\begin{equation}
H^2=\frac{\Lambda_{4}}{3}+\bigg(\frac{8\pi}{3M^{2}_{4}}\bigg)\rho+
\bigg(\frac{4\pi}{3M^{3}_{5}}\bigg)\rho^{2}+\frac{{\cal{E}}_{0}}{a^{4}}
\end{equation}
Where $M_{4}$ and $M_{5}$ are four and five dimensional fundamental
scales respectively and $\Lambda_{4}$ is the effective cosmological
constant on the brane. The last term in equation (3) is called the
dark radiation term and ${\cal{E}}_{0}$ is an integration constant.
The relation between four and five dimensional fundamental scales is
\begin{equation}
M_{4}=\sqrt{\frac{3}{4\pi}}\bigg(\frac{M^{2}_{5}}{\sqrt{\lambda}}\bigg)M_{5}
\end{equation}
where $\lambda$ is the brane tension. We now suppose that the
initial singularity that leads to RS II geometry afterwards, is
smeared due to noncommutativity of the spacetime. A newly proposed
model for the similar scenario in the usual 4D universe suggests
that one could split the energy density on any hypersurface as [6]
\begin{equation}
\rho=\rho_{0}e^{-|\tau|^{2}/4\theta}e^{-|\vec{{X}}|^{2}/4\theta}
\end{equation}
where $R^{2}=\tau^{2}+|\vec{X}|^{2}$ and $\tau=it$ is the Euclidean
time. Note that we suppose that the universe enters the RS II
geometry immediately after the initial smeared singularity which is
a reasonable assumption ( for instance, from a M-theory perspective
of the cyclic universe this assumption seems to be reliable, see
Ref. [10]). From one hypersurface to another, the
$\vec{X}$-dependent part of $\rho$ does not change, so it can be
included into $\rho_{0}$. If we neglect the dark radiation term
(which is reasonable during inflation as it is vanishing really
fast\footnote{ But note that this term is important when one treats
the perturbations on the brane. As has been shown in Ref. [11], on
large scales this term slightly suppresses the radiation density
perturbations at late times. In a kinetic era, this suppression is
much stronger and drives the density perturbations to zero.)}) and
also the brane cosmological constant, the Friedmann equation (3) can
be rewritten as
\begin{equation}
H^{2}=\frac{8\pi}{3M^{2}_{4}}\rho\bigg[1+\frac{\rho}{2\lambda}\bigg]
\end{equation}
Using equation (5), this Friedmann equation in noncommutative space
could be rewritten as follows
\begin{equation}
\bigg(\frac{\dot{a}}{a}\bigg)^{2}=\frac{8\pi}{3M^{2}_{4}}\rho_{0}e^{-t^{2}/4\theta}
\bigg[1+\frac{\rho_{0}e^{-t^{2}/4\theta}}{2\lambda}\bigg].
\end{equation}
This equation can be solved for $a(t)$ to obtain
\newpage

$$a \left( t \right) ={\cal{H}} \bigg( \Big[\frac{1}{4}\,{\frac {\rho_{0}
-2\,\sqrt {2}\theta\,{\lambda}^{3/2}\sqrt
{\frac{8\pi}{3M^{2}_{4}}}}{\rho_{0}}}\Big]\,,\,\,\Big[\frac{1}{2}\Big]\,,\,\,\frac{1}{8}\,{\frac
{\sqrt {2}\sqrt {\frac{8\pi}{3M^{2}_{4}}} \left[  \left(
4\,\rho_{0}+4\,\lambda \right) \theta+t \rho_{0}
 \right] ^{2}}{\theta\,\sqrt {\lambda}\rho_{0}}} \bigg)\times$$
$${\exp\bigg\{-\frac{1}{16}\,{\frac {
 \left[ \left( 8\,\rho_{0}+8\,\lambda \right) \theta+t \rho_{0}\right] \sqrt {2}
\sqrt {\frac{8\pi}{3M^{2}_{4}}}t }{\theta\,\sqrt
{\lambda}}}}\bigg\}\,\,\,+$$
$$\left[ \left( 4 \,\rho_{0}+4\,\lambda \right) \theta+t \rho_{0}
\right]
 {\exp\bigg\{-\frac{1}{16}\,{\frac {
\left[
 \left( 8\,\rho_{0}+8\,\lambda \right) \theta+t \rho_{0}\right] \sqrt {2}\sqrt {\frac{8\pi}{3M^{2}_{4}}}t }
{\theta\,\sqrt {\lambda}}}}\bigg\}\times$$
\begin{equation}
{\cal{H}} \bigg( \Big[\frac{1}{4}\,{\frac {3\,\rho_{0}- 2\,\sqrt
{2}\theta\,{\lambda}^{3/2}\sqrt
{\frac{8\pi}{3M^{2}_{4}}}}{\rho_{0}}}\Big]\,,\,\,\Big[\frac{3}{2}\Big]\,,\,\,\frac{1}{8}\,{\frac
{ \sqrt {2}\sqrt {\frac{8\pi}{3M^{2}_{4}}} \left[  \left(
4\,\rho_{0}+4\,\lambda \right) \theta+t \rho_{0}
 \right] ^{2}}{\theta\,\sqrt {\lambda}\rho_{0}}} \bigg)\,,
\end{equation}
where ${\cal{H}}$ shows the Hypergeometric function of the
arguments. To see the cosmological dynamics of the model, we plot
the evolution of the scale factor and the Hubble parameter in
figures $1$ and $2$. As figure $1$ shows, this noncommutative model
naturally gives an inflation era without consulting to any axillary
inflaton field. On the other hand, due to smeared picture adopted in
this noncommutative framework, there is no initial singularity in
this setup.
\begin{figure}[htp]
\begin{center}
\includegraphics{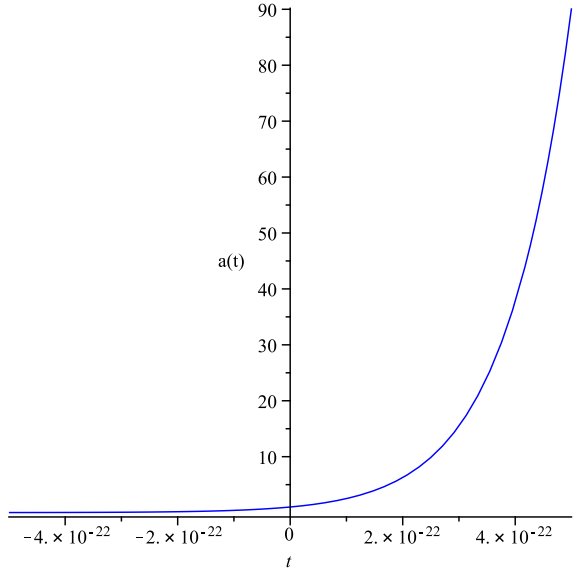}
\end{center}
\vspace{7.5 cm} \caption{\small {Evolution of the scale factor in
noncommutative Randall-Sundrum II geometry. There is an inflationary
era without recourse to any scalar or vector fields. The model
avoids also the initial singularity.}}
\end{figure}

\begin{figure}[htp]
\begin{center}
\includegraphics{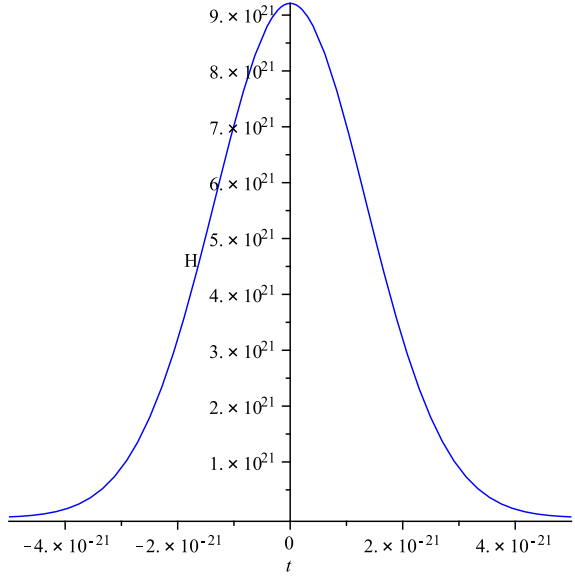}
\end{center}
\vspace{6 cm} \caption{\small {Evolution of the Hubble parameter in
noncommutative Randall-Sundrum II geometry}}

\end{figure}
The number of e-folds in this model will be given by
$$
N=\int^{t_{f}}_{t_{i}}H dt\simeq \frac{8}{3}\pi \,{\it \rho_{0}}\,
\bigg[ \sqrt {\pi \theta}\,\,\,{\rm erf}
 \Big(\frac{1}{2}\,{\frac {t_{f}}{\sqrt {\theta}}} \Big) +\frac{1}{2}\,\sqrt {2\pi
 \theta}\,\,\,
{\rm erf} \Big( \frac{1}{2}\,{\frac {\sqrt {2}t_{f}}{ \sqrt
{\theta}}} \Big) {\lambda}^{-1} \bigg] {M_{4}}^{-2}$$
\begin{equation}
- \frac{8}{3}\,\pi \,{\it \rho_{0}}\, \bigg[ \sqrt {\pi
\theta}\,\,\, {\rm erf} \Big( \frac{1}{2}\,{\frac {t_{i}}{\sqrt
{\theta}}} \Big) +\frac{1}{2}\,\sqrt {2\pi \theta}\,\,\, {\rm erf}
\Big( \frac{1}{2}\,{\frac {\sqrt {2}t_{i}}{ \sqrt {\theta}}} \Big)
{\lambda}^{-1} \bigg] {M_{4}}^{-2}.
\end{equation}
By expanding the error functions in equation (9) in series, the
number of e-folds (supposing that the universe enters the
inflationary phase immediately after the big bang,\, that is,\,
$t_{i}=0$ and $t_{f}=t$) will be given by
\begin{equation}
N\simeq \frac{8}{3}\,\pi \,{\it \rho_{0}}\, \bigg[
t-\frac{1}{12}\,{\frac {{t}^{3}}{\sqrt {\pi }{
\theta}^{\frac{3}{2}}}}+{\frac {1}{160}}\,{\frac {{t}^{5}}{\sqrt
{\pi }{\theta }^{\frac{5}{2}}}}+\frac{1}{2}\, \Big(
2\,t-\frac{1}{6}\,{\frac {\sqrt {2}{t}^{3}}{\sqrt {\pi
}{\theta}^{\frac{3}{2}}}}+\frac{1}{40}\,{\frac {\sqrt
{2}{t}^{5}}{\sqrt {\pi }{\theta} ^{\frac{5}{2}}}} \Big)
{\lambda}^{-1} \bigg] {M_{4}}^{-2}.
\end{equation}
We plot this relation as a function of time in figure (3). It is
obvious from this figure that if $\rho_{0}$ is suitably large, we
will get sufficient amount of inflation in this scenario.
\begin{figure}[htp]
\begin{center}
\includegraphics{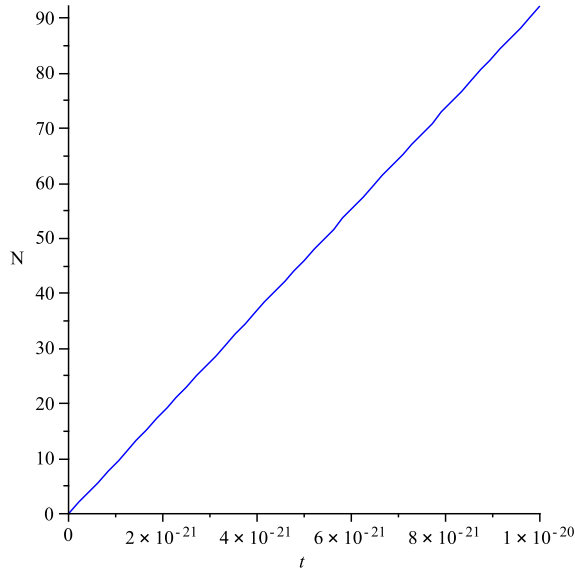}
\end{center}
\vspace{7 cm} \caption{\small {Number of e-folds as a function of
time. If the amount of $\frac{\rho_{0}}{\lambda}$ in equation (10)
is suitably large, we will get sufficient amount of inflation. We
have set $\theta=10^{-20}$ and $\frac{\rho_{0}}{\lambda}=10^{16}$
with $M_{4}=1$. }}
\end{figure}
Now, using equation (10) and solving for $\rho_{0}$, we find
\begin{equation}
 \rho_{0}=45\,\sqrt {\theta}\lambda\,{M_{4}}^{2} \bigg[ 2\,{\pi
}^{\frac{3}{2}} \theta\,\lambda\,t\,\,\,{\rm erf}
\Big(\frac{1}{2}\,{\frac {t}{\sqrt {\theta}}}
 \Big) +4\,\pi \,{\theta}^{\frac{3}{2}}\lambda\,{e^{-\frac{1}{4}\,{\frac {{t}^{2}}{
\theta}}}}+{\pi }^{\frac{3}{2}}\theta\,\sqrt {2}t\,\,\, {\rm erf}
\Big( \frac{1}{2}\,{\frac {\sqrt {2}t}{\sqrt {\theta}}} \Big)
+2\,\pi \,{\theta}^{\frac{3}{2}}{e^{-\frac{1}{2} \,{\frac
{{t}^{2}}{\theta}}}} \bigg]^{-1}.
\end{equation}
Usually the number of e-folds required to solve problems of standard
cosmology is $N\simeq 60-70$. If we assume that the value of
\,$\theta$\, to be of the order of $10^{-20}$, the value of
$\frac{\rho_{0}}{\lambda}$ required for a successful inflation with
$N=60$ is $\frac{\rho_{0}}{\lambda}\sim 10^{16}$ where we have set
$M_{4}=1$. We note that $\frac{\rho_{0}}{\lambda}$ obtained in this
way is a fine-tuned value. The value of \,$\theta$\, can be
estimated for instance by the noncommutative correction to the
planets perihelion precession of the solar system [12] (see also
[3]). Another point we stress here is that Rinaldi has pointed in
Ref. [6] that $\frac{\rho_{0}}{\lambda}$ may play the role of a
cosmological constant after the inflationary phase. Actually this is
not the case since $\frac{\rho_{0}}{\lambda}$ has not correct
equation of state to be dark energy.

To be a realistic model of the early universe and also to test
whether or not our model is consistent with recent observational
data, a scale invariant spectrum of scalar perturbations should be
generated after inflation. We define the slow-roll parameters as
usual
$$\epsilon\equiv\frac{{M_{4}}^{2}}{4\pi}\bigg(\frac{H'}{H}\bigg)^{2},$$
\begin{equation}
\eta\equiv\frac{{M_{4}}^{2}}{4\pi}\bigg(\frac{H''}{H}\bigg)
\end{equation}
These slow-roll parameters as a function of cosmic time are given in
appendix $1$. We assume that as usual the scalar spectral index is
given by the
\begin{equation}
n_{s}-1\simeq -6\epsilon+2\eta.
\end{equation}
This assumption will be justified shortly. To match the
observational data, $n_{s}$ should be around unity at the end of
inflation. This guaranties the generation of scale invariant scalar
perturbations. Figure $4$ shows variation of $n_{s}$ versus the
cosmic time. In plotting this figure we have used the same values of
parameters as have been used to plot figure $3$. As one can see from
this figure, it is possible essentially to have scale invariant
scalar spectrum in this model. However, we stress that in order to
study the power spectrum in our model, a more thorough analysis of
generation of density perturbations should be done, taking into
account the dark radiation term since this term plays a crucial role
in perturbations. Especially the relation (13) needs to be
reformulated in this non-commutative framework. These issues are
under investigation by the authors.\\

\begin{figure}[htp]
\begin{center}
\includegraphics{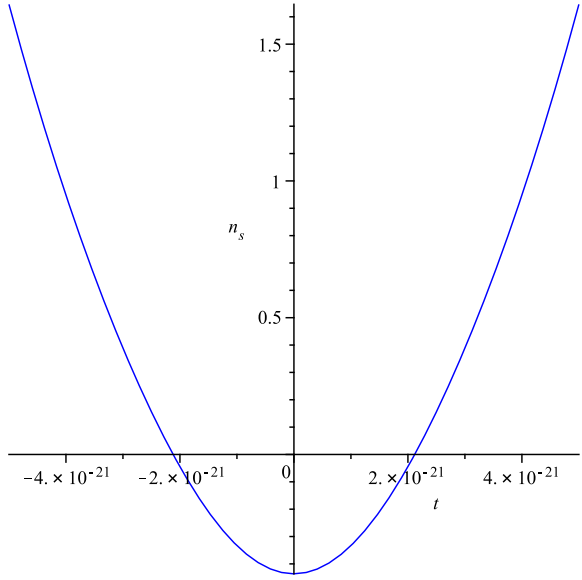}
\end{center}
\vspace{7 cm} \caption{\small {Variation of the scalar spectral
index versus the cosmic time. The spectral index approaches the
Harrison-Zel'dovic spectrum at the end of inflation. The parameters
used to plot this figure are the same as previous figures. The
spectral index is exactly one at $t=\pm4.021168857\times10^{-21}$ }}
\end{figure}
Finally, two points should be explained here: Firstly, one might
think that this model has the potential to be able to solve the
flatness problem from an accelerated expansion. We note however that
this is not actually the case since as long as this model want to
address the singularity problem, one needs to consider $t\rightarrow
-\infty$, where $\rho$ (defined in Eq. (5)) is exponentially small.
If there were large spatial curvature when $t\rightarrow -\infty$,
the spatial curvature will dominate the universe quickly. Secondly,
as one can read from Fig. $2$, with our choice of parameters, the
end of the inflation era takes place around $4\times10^{-21}$. In
order to have a scale invariant spectrum of scalar perturbations,
$n_{s}$ should be around unity at this time. From figure $4$ one can
see that this is indeed the case. The scalar spectral index is
exactly one at the time $t=\pm4.021168857\times10^{-21}$. $n_{s}$
changes from negative values at $t=0$ to around unity at the end of
inflationary era.

\section{Summary}
In this letter, by adopting the smeared coherent state picture of
spacetime noncommutativity, we generalized the Randall-Sundrum II
braneworld inflation to noncommutative spaces. This model realizes
an inflationary, bouncing solution without recourse to any axillary
scalar or vector fields. Due to noncommutative structure of the very
spacetime which admits the existence of a fundamental length scale,
there is no initial singularity in this model. Note that we supposed
that the universe enters the RS II geometry immediately after the
initial smeared singularity which is a reasonable assumption for
instance from a M-theory perspective of the cyclic universe. There
is a parameter, $\frac{\rho_{0}}{\lambda}$, in this model that has
the potential to play important roles in the inflation era: by
taking the number of e-folds to be $N\simeq60$, and setting the
noncommutativity parameter to be $\theta\sim 10^{-20}$, the value of
$\frac{\rho_{0}}{\lambda}$ required for a successful inflation is
$\frac{\rho_{0}}{\lambda}\sim 10^{16}$. By treating the scalar
perturbations in this setup, we have shown that it is possible
essentially to have scale invariant scalar perturbations in this
framework. From another viewpoint, $\rho_{0}$ contains a
space-dependent part of $e^{-|\vec{{X}}|^{2}/4\theta}$ that
essentially breaks the homogeneity on the successive hypersurfaces.
This may open new windows on the issue of cosmological
perturbations. A more thorough analysis of perturbations on the
brane is therefore required to justify the successes of this model.\\

\begin{center}
\bf{Appendix 1}\\ \vspace{0.5cm} {\bf Slow-roll Parameters}
\end{center}
The slow-roll parameters defined in equation (12) are given by
$$\epsilon=\frac{{M_{4}}^{2}}{4\pi}{\frac {9}{64}}\, \Bigg[
-\frac{4}{3}\,\pi \,{\rho_{0}}\,t{{\rm e}^{-\frac{1}{4}\,{ \frac
{{t}^{2}}{\theta}}}} \left( 1+{{\rm e}^{-\frac{1}{4}\,{\frac
{{t}^{2}}{ \theta}}}}{\lambda}^{-1} \right)
{\theta}^{-1}{M_{4}}^{-2}-\frac{4}{3}\,\pi \,{ \rho_{0}}\,  {{\rm
e}^{-\frac{1}{2}\,{\frac {{t}^{2}}{\theta}}}}
t{\theta}^{-1}{\lambda}^{-1}{M_{4}}^{-2}\Bigg]^{2}\times $$
\begin{equation}
\Bigg[{M_{4}}^{4}{\pi }^{-2} {\rho_{0}}^{-2} {{\rm
e}^{-\frac{1}{2}\,{\frac {{t}^{2}}{\theta}}}}
 \left( 1+{{\rm e}^{-\frac{1}{4}\,{\frac {{t}^{2}}{\theta}}}}{
\lambda}^{-1} \right) ^{-2}\Bigg]\,
\end{equation}
and
$$\eta=\frac{{M_{4}}^{2}}{4\pi}\frac{3}{8}\, \Bigg[ -\frac{4}{3}\,\pi
\,{\rho_{0}}\,{{\rm e}^{-\frac{1}{4}\,{\frac {{t}^{2}}{ \theta}}}}
\left( 1+{{\rm e}^{-\frac{1}{4}\,{\frac {{t}^{2}}{\theta}}}}{
\lambda}^{-1} \right) {\theta}^{-1}{M_{4}}^{-2}+\frac{2}{3}\,\pi
\,{\rho_{0}}\,{t}^{ 2}{{\rm e}^{-\frac{1}{4}\,{\frac
{{t}^{2}}{\theta}}}} \left( 1+{{\rm e}^{-\frac{1}{4} \,{\frac
{{t}^{2}}{\theta}}}}{\lambda}^{-1} \right) {\theta}^{-2}{M}^{
-2}$$$$+2\,\pi \,{\rho_{0}}\,{t}^{2} {{\rm e}^{-\frac{1}{2}\,{\frac
{{t}^{2}}
{\theta}}}}{\theta}^{-2}{\lambda}^{-1}{M_{4}}^{-2}-\frac{4}{3}\,\pi
\,{\rho_{0}}\, {{\rm e}^{-\frac{1}{2}\,{\frac {{t}^{2}}{\theta}}}}
{\theta}^{-1}{\lambda}^{-1}{M}^{-2} \Bigg]\times$$
\begin{equation}
\Bigg[{M_{4}}^{2}{\pi } ^{-1}{\rho_{0}}^{-1}{{\rm
e}^{\frac{1}{4}\,{\frac {{t}^{2}}{\theta}}} } \left( 1+{{\rm
e}^{-\frac{1}{4}\,{\frac {{t}^{2}}{\theta}}}}{ \lambda}^{-1} \right)
^{-1}\Bigg]\,.
\end{equation}

\end{document}